*electronics*

MDPI

*Article*

# A Method for Evaluating Chimeric Synchronization of Coupled Oscillators and Its Application for Creating a Neural Network Information Converter

**Andrei Velichko**

Institute of Physics and Technology, Petrozavodsk State University, 31 Lenina str., Petrozavodsk 185910, Russia; velichko@petrsu.ru; Tel.: +7-8142-63-5773

**Abstract:** This paper presents a new method for evaluating the synchronization of quasi-periodic oscillations of two oscillators, termed "chimeric synchronization". The family of metrics is proposed to create a neural network information converter based on a network of pulsed oscillators. In addition to transforming input information from digital to analogue, the converter can perform information processing after training the network by selecting control parameters. In the proposed neural network scheme, the data arrives at the input layer in the form of current levels of the oscillators and is converted into a set of non-repeating states of the chimeric synchronization of the output oscillator. By modelling a thermally coupled $VO_2$-oscillator circuit, the network setup is demonstrated through the selection of coupling strength, power supply levels, and the synchronization efficiency parameter. The distribution of solutions depending on the operating mode of the oscillators, sub-threshold mode, or generation mode are revealed. Technological approaches for the implementation of a neural network information converter are proposed, and examples of its application for image filtering are demonstrated. The proposed method helps to significantly expand the capabilities of neuromorphic and logical devices based on synchronization effects.

## 1. Introduction

Artificial neural networks are actively used in image and speech recognition applications [1,2], as well as in computer calculations [3] and data coding [4]. The functional importance of synchronization in information processing has stimulated the development of neural network models with oscillatory dynamics and neuromorphic algorithms based on synchronization effect.

In 2002, Kuramoto and Battogtokh reported that arrays of non-locally related oscillators could spontaneously divide into synchronized and desynchronized subpopulations [5]. This amazing discovery challenged the previous belief that the connected identical oscillators are either synchronized or will work incoherently and chaotically. Since the network had a hybrid nature uniting both coherent and non-coherent parts, it was proposed to call such states chimera, because of their resemblance to mythological Greek animals, assembled from incomparable parts [6]. Recent studies have demonstrated that chimera states are not limited to phase oscillators, but can be found in a wide variety of different systems and are observed in space-time dynamics. It is worth mentioning the studies on chimeric states in networks of Kuramoto phase oscillators [7], and leaky integrate-and-fire [8] FitzHugh-Nagumo [9] and Hindmarsh-Rose [10] models. In a previous study [11], chimera states are classified into stationary, turbulent, and breathing patterns, which, in turn, are divided into subclasses. In addition, a separation of amplitude and phase chimeras is identified [12,13]. Such a wide range of gradations is due to the diversity of the spatial and temporal network dynamics.

Consequently, the term "chimera states" has been well established in the literature. It characterizes an array of connected oscillators, where some of the oscillators operate in synchronous mode, and other oscillators function in asynchronous mode [14]. However, alternative terms of this phenomenon are found in the literature, for example, in a previous study [15], this behavior is called cluster synchronization.

In this paper, the term "chimeric synchronization" is used to refer to the synchronization effect and to describe the coexistence of synchronous and asynchronous states of coupled oscillators in the time domain. Chimeric synchronization is the phenomenon of synchronization of quasi-periodic oscillations [16–21]. Quasi-periodic oscillations can be observed when crossing the boundaries of the Arnold tongue, describing high order synchronization, during a transition from the synchronous to asynchronous state. A number of authors [16–21] have studied the synchronization of quasiperiodic oscillations; however, no specific methodology for evaluating this phenomenon for technical applications exists.

To determine the type of chimera state and for estimating the synchronization of quasi-periodic oscillations, the degree of synchronization is evaluated by Lyapunov spectra [12,22] or by calculating the complex order parameter [23]. However, these indicators do not contain information about the complex structure of synchronization, which, in the case of chimeric synchronization, consists of a set of different individual synchronous patterns.

Therefore, the development of methods for assessing chimeric synchronization and its application in practice is an important task. In addition, the author makes an assumption about the relationship of the formation of chimeric states and the phenomenon of chimeric synchronization, which is discussed in the paper results.

A class of oscillatory neural networks (ONN) can be identified, where the basic elements are relaxation oscillators that generate sequences of pulses (spikes), and ONN can encode information at pulses repetition rate. Such ONN are interesting due to simplicity of hardware implementation, as developed micro- and nano-electronic autogenerators ensure network compactness and energy efficiency. In addition, a pulsed-type ONN, where the periodic oscillation spectrum has a multi-frequency character, has a special mode of harmonics synchronization, or in other words, has a high order synchronization effect [24–26]. This effect has been demonstrated experimentally using the example of thermally coupled $VO_2$ oscillators [24]. In relaxation generators with elements based on vanadium dioxide film, oscillations are initiated by the electric switching effect caused by the metal-insulator phase transition [27].

Oscillators based on $VO_2$ structures are chosen as ONN elements due to the high speed of electrical switching ($\sim 10$ ns) [28], high degree of nanoscale in manufacturing [29], and, most importantly, the presence of a significant thermal coupling effect that simplifies the ONN layout and circuitry of galvanically isolated oscillators [24]. The thermal coupling effect, having a local nature, makes the network resemble a cellular neural network (CNN) [3,30] and opens up 3D integration possibilities. For these reasons, $VO_2$-oscillators are actively used in prototyping of neuro-oscillators for the tasks of cognitive technologies.

The high-order synchronization of two oscillators can be characterized by a family of metrics—the ratio of subharmonics (SHR) and synchronization effectiveness $\eta$ [26]. These metrics are used to create a neuromorphic device for pattern recognition. Synchronization is measured relative to the reference oscillator, which has a constant pulse generation frequency; this makes it possible to compare the synchronization of all oscillators in the network with each other. Although the system dynamics during synchronization can be described only by SHR and $\eta$ value, there are a number of examples where oscillograms contain sections with different synchronization, termed chimeric synchronization in this paper. Chimeric synchronization appear quite often in the model, however, it does not affect operation of the device and belongs to an asynchronous type of oscillation. For example, Figure 1 shows the current pulses of two oscillators with numbers 0 and 1, with alternating synchronization patterns defined by two parameters $SHR^1_{0,1}$ = 4:3 and $SHR^2_{0,1}$ = 9:7. The dotted lines mark the synchronization moments (phase locked) of the individual pulses.

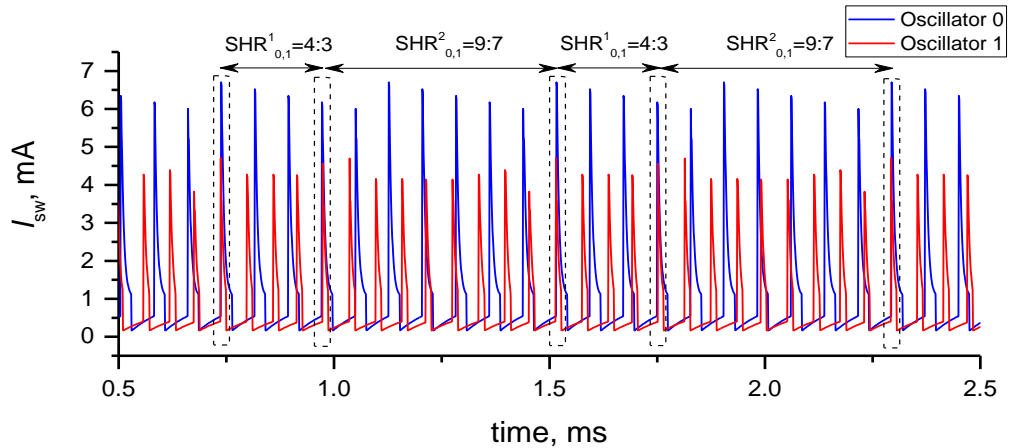

**Figure 1.** An example of a chimeric synchronization of oscillators with an alternating synchronization pattern of current pulses in an oscillogram. The dotted line marks synchronization moments of individual pulses. $I_{sw}$ is the oscillator current.

Such an alternation of the synchronization pattern is a stable state in time and is characterized by a periodic jump of the phase of synchronous pulses.

In this paper, we describe the method for classification of chimeric synchronization and use it to analyze the dynamics of six thermally coupled $VO_2$ oscillators as applied to the implementation of a neural network converter.

## 2. Materials and Methods

We describe the classification method of chimeric synchronization at the beginning of this section, followed by a description of the single-oscillator circuit, ONN structure, formulation of a search problem and network training technique, and technical aspects of thermal coupling organization.

*2.1. Chimeric Synchronization Classification*

The method of determining the family of metrics of high-order synchronous states (SHR and μ) is described in detail in a previous study [26]. The method of classification of chimeric synchronization is based on its basis and is not significantly different.

At first, the analog oscillogram of oscillations is represented in the form of the corresponding array, called LE, which stores information on the position of the current pulse leading edges (see Figure 2).

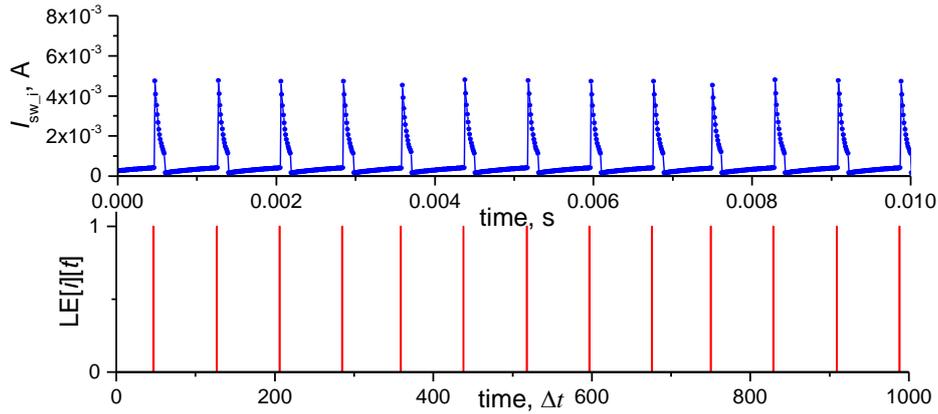

**Figure 2.** Oscillogram of oscillator current and the corresponding array of positions of the leading edges of the current pulse LE[*i*][*t*]. (where *i* is the oscillator number, time *t*=*n*·Δ*t*, n is number of the calculation steps of the model oscillogram, Δ*t* is the calculation time interval).

Then, for two arbitrary oscillators *i* and *j*, between which it is necessary to determine synchronization, the arrays LE[*i*][*t*] and LE[*j*][*t*] are compared (see. Figure 3). The distance between the two nearest phase-locked pulses is denoted as $T^z_s$, the period of synchronization (where *z* is a conditional number of periods $T_s$). Pulses are considered phase-locked if the distance between them does not exceed 4Δ*t*.

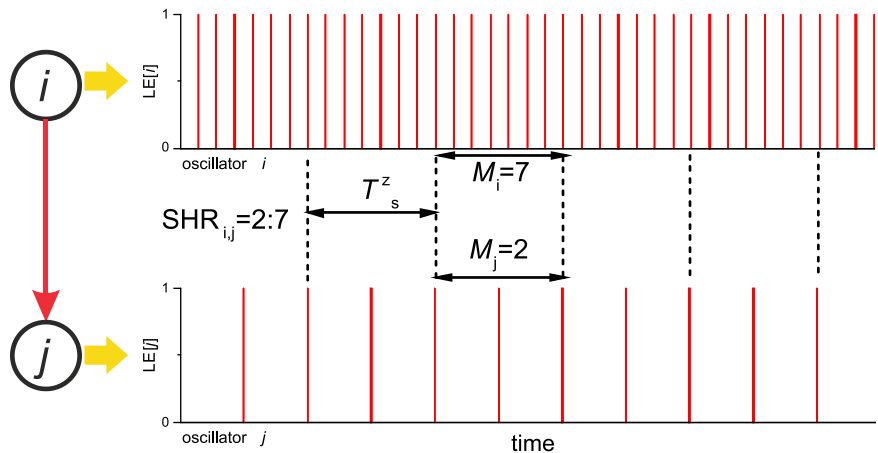

**Figure 3.** Arrays LE[*i*][*t*] and LE[*j*][*t*] for two oscillators.

The value of sub-harmonic ratio between the oscillator *i* and the oscillator *j*, called $SHR_{i,j}$, may be estimated using a phase-locking method:

$$\text{SHR}_{i,j} = M_j : M_i \quad (1)$$

where $M_i$ and $M_j$ are numbers of signal periods falling into the synchronization periods $T^z_s$ of two oscillators. Equation (1) defines the pattern of high order synchronization of two signals. As the notation "SHR" defines, the formula can determine the ratio of subharmonic numbers in the signal spectra at a common synchronization frequency [26].

In general, especially when a system behaves erratically, synchronization periods differ and spread in $T^z_s \neq T^{z+1}_s$ and the values of $M_i$ and $M_j$ may change within one oscillogram (see Figure 4).

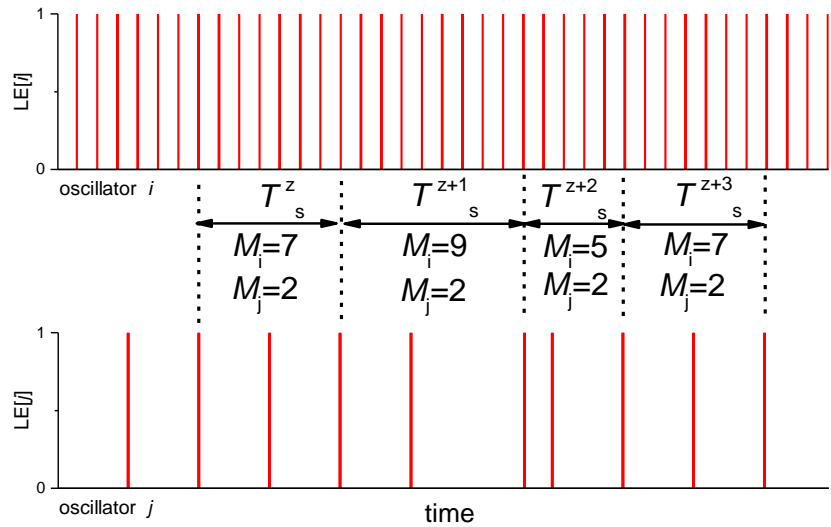

**Figure 4.** Arrays LE[*i*][*t*] and LE[*j*][*t*] for two oscillators with non-constant period of synchronization $T^z{}_s$.

Various values of synchronizations $SHR_{i,j}$ may occur within one oscillogram. To determine the distribution of the $SHR_{i,j}$, it is necessary to find the occurrence probabilities $P(M_j : M_i)$ for each pair $(M_i : M_j)$ that are present in the whole oscillogram. To find the probabilities $P(M_j : M_i)$, we can count how many times $NP(M_j : M_i)$ the given pair appeared within the whole oscillogram of the oscillator *i*, multiply this by the number of periods in it ($M_i$), and divide this by the total number of all oscillation periods in the given signal ($N_j$). Thus, for $P(M_j : M_i)$ we obtain:

$$P(M_j : M_i) = 100\% \cdot NP(M_j : M_i) \cdot M_i / N_i \qquad (2)$$

where $N_i$ is the total number of periods in the oscillogram of oscillator *i*.

Therefore, each synchronization value $SHR_{i,j}$ will correspond to the probability of its detection $P(M_j : M_i)$, expressed as a percentage.

It is convenient to present the probabilities $P(M_j : M_i)$ as a histogram, where the values are positioned in the descending order of the magnitude *P*. For example, for the oscillogram section in Figure 4, the following histogram can be performed:

The histogram in Figure 5 is calculated by Equation (2), when the pairs occur a number of times $NP(2:7)=2$, $NP(2:9)=1$, $NP(2:5)=1$, and the total number of periods is $N_i=28$ (in real calculations, $N_i$ was in the range of 1000–3000 for greater accuracy [26]).

In a model experiment, the shapes of the distribution oscillograms $P(M_j : M_i)$ can differ significantly from each other. For example, Figure 6 presents the main histogram variants occurring during signal processing (oscillator *i* corresponds to the reference oscillator with a constant frequency).

The histogram in Figure 6a corresponds to the case of an absolutely synchronized signal with high order synchronization $SHR_{i,j} = 11:8$. The spectrum of oscillator *j* has a line character, and the phase diagram corresponds to a single high order synchronization limit cycle. The cases in Figure 6b,c have a set of different $SHR_{i,j}$ and correspond to chimeric synchronization states.

A chimeric index, called CH, can be introduced, consisting of the first three values of the synchronization value $SHR_{i,j}$:

$$CH_{i,j} = (SHR^1{}_{i,j}, SHR^2{}_{i,j}, SHR^3{}_{i,j}) \qquad (3)$$

Accordingly, in Figure 6, each histogram displays its values: (a) $CH_{i,j}$ = (11:8), (b) $CH_{i,j}$ = (3:2 7:5), and (c) $CH_{i,j}$ = (11:7 24:17 8:5). For Figure 6a, the values of $CH_{i,j}$ and $SHR_{i,j}$ are the same.

For CH, the parameter of synchronization effectiveness $\eta$ is defined as the sum of P ($M_j, M_i$) for the first three values in the histogram:

$$\eta = \sum_{k=1}^{3} P^k (M_j : M_i) \quad (4)$$

where *k* is the sequence number P($M_j$, $M_i$) in the histogram.

Therefore, for each histogram (see Figure 6), we define the effectiveness: (a) $\eta$ = 97.8%, (b) $\eta$ = 99.7%, and (c) $\eta$ = 74.8%.

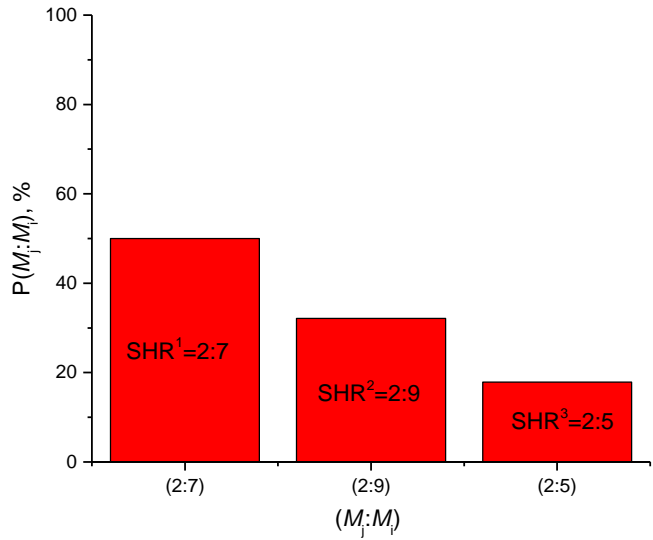

**Figure 5.** Histogram of probabilities distribution P($M_j$, $M_i$), calculated by using Equation (2) for signals of LE, shown in Figure 5.

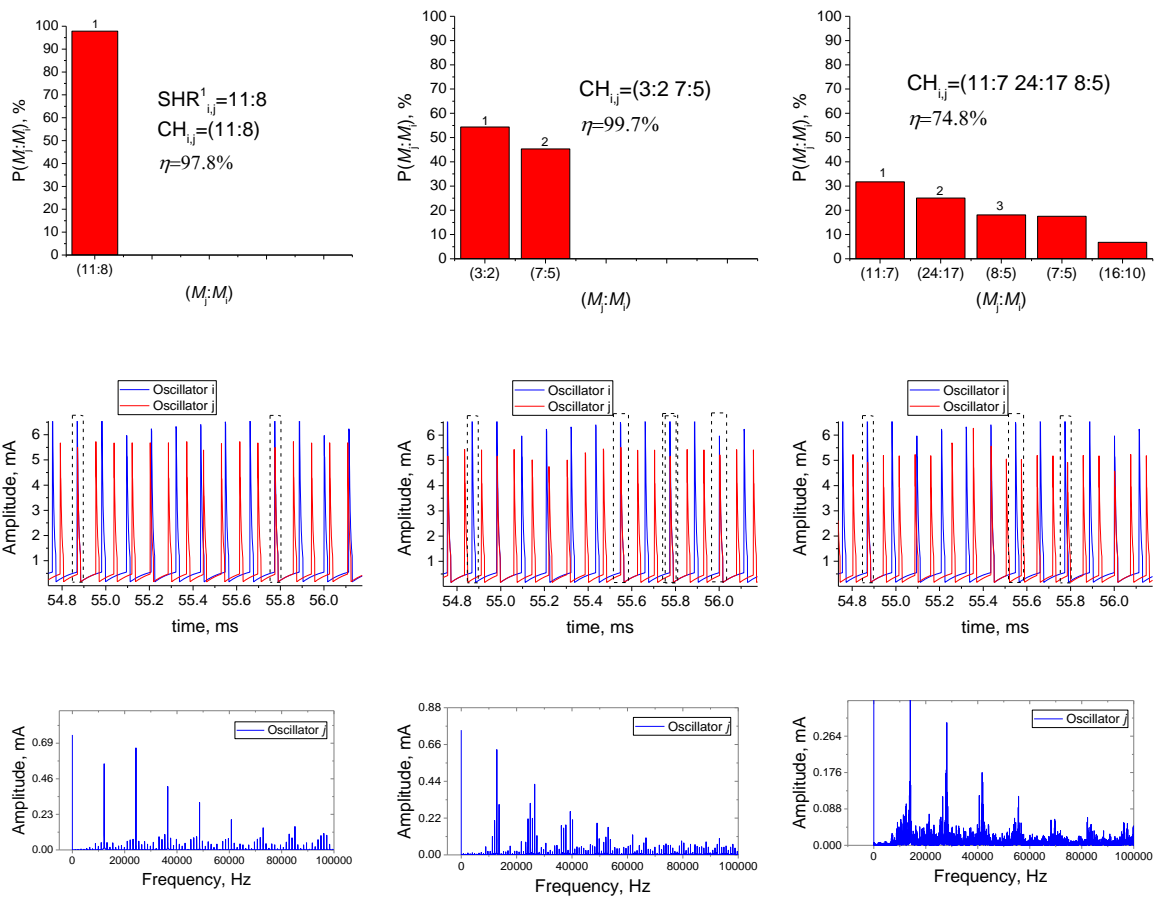

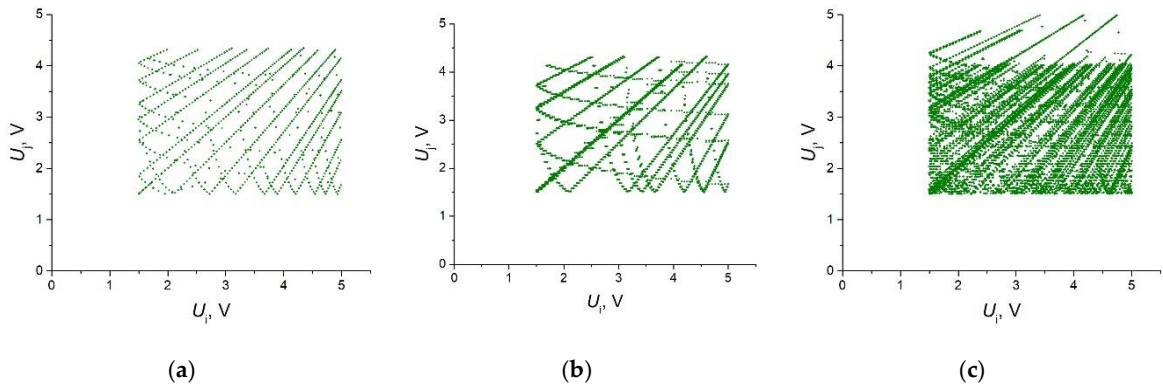

**Figure 6.** Synchronization characteristics of two oscillators: the main types of histograms occurring during signal processing and the corresponding oscillograms, spectra, and phase diagrams. (**a**) $CH_{i,j}$ = (11:8), (**b**) $CH_{i,j}$ = (3:2 7:5), and (**c**) $CH_{i,j}$ = (11:7 24:17 8:5). The dotted line indicates the synchronous peaks of the two oscillators.

The new family of metrics ($CH_{i,j}$, $\eta$) allows sufficient determination of the synchronization states of two oscillators, and classification of chimeric synchronization. Depending on the task, for example, the network training for data coding and pattern recognition, the problem of the presence or absence of synchronization can be solved by formally setting the synchronization effectiveness threshold $\eta_{th}$, so

$$\text{signals are} \begin{cases} \text{synchronized, if } \eta \geq \eta_{th} \\ \text{not synchronized, if } \eta < \eta_{th} \end{cases} \qquad (5)$$

In the majority of cases, we set $\eta_{th}$ = 90%, meaning the signals are synchronized if 90% of their durability has a certain synchronization pattern or a set of patterns of the chimeric synchronization. For the network training, this parameter can be selected within a selected range, and it is one of the important parameters of the network adjustment [31].

Let us discuss the reasons for the introduction of the concept of CH index chimera synchronization. For signals in Figure 6b, the synchronization is $CH_{i,j}$ = (3:2 7:5) and $\eta$ = 99.7%, thus, the signals are clearly synchronized ($\eta$ > 90%) and have two patterns of synchronization. The spectrum of oscillator *j* has a linear character, and the phase diagram of voltages on oscillators has a complex, but not chaotic, attractor, most likely consisting of two limit cycles. If the technique [26] and only the concept of basic synchronization $SHR_{i,j}$ are applied, then the family of metrics would look like $SHR_{i,j}$ = 3:2 and $\eta$ = 54%. As a result, oscillators would be defined as not synchronized, since $\eta$ <90%. However, an accurate calculation of the chimeric synchronization value using the proposed metric allows more accurate and complete characterization of the synchronization state. In addition, such a metric can significantly expand the capabilities of neuromorphic and logical devices that operate on the synchronization effect.

For signals in Figure 6c, the synchronization is $CH_{i,j}$ = (11:7 24:17 8:5) and $\eta$ = 74.8%, so the signal is weakly synchronized, and at $\eta_{th}$ = 90%, it is not formally synchronized, as in Equation (5). This is confirmed by the type of phase trajectory that fills the entire phase space. In addition, the oscillator *j* spectrum is wide, contains many harmonics, and is close to the noise spectrum by its nature.

The main technical problem we faced was the problem of defining the synchronization between the reference oscillator No.0 and the oscillator of the output layer No.5 characterized by the values $SHR_{0,5}$ and $CH_{0,5}$:

$$\begin{aligned} SHR_{0,5} &= M_0 : M_5 \sim M_0 / M_5 \\ CH_{0,5} &= (SHR^1_{0,5}, SHR^2_{0,5}, SHR^3_{0,5}) \end{aligned} \qquad (6)$$

Two parameters $CH_{0,5}$ and $\eta$ are used as the main metrics for evaluation the degree of the two oscillators' synchronization and are applied in the algorithm of ONN training.

Current oscillograms $I_{sw}(t)$ of oscillators No.0–5 were calculated simultaneously and contained ~250,000 points with time interval $\Delta t = 1$ µs. Then, the oscillograms were automatically processed.

*2.2. Method of Chimeric Synchronization Color Mapping*

To represent the value of chimeric synchronization $CH_{i,j}$, we chose the RGB color display method, since the value of $CH_{i,j}$ contains three components, as in Equation (6). Each color component is represented by the following algorithm:

$$CH_{i,j} \rightarrow RGB(red, green, blue)$$
$$CH_{i,j}(SHR^1, SHR^2, SHR^3) \rightarrow RGB(\lambda^1, \lambda^2, \lambda^3)$$

$$\lambda^k = \begin{cases} 100+(SHR^k/SHR_{max}) \cdot 100, & \text{if } SHR^k \geq 1 \\ 100-((1/SHR^k)/SHR_{max})) \cdot 100, & \text{if } SHR^k < 1 \end{cases} \quad (7)$$
$$k = 1 \ldots 3$$

where $SHR_{max}$ is the maximum value of SHR on the graph, and k is an index denoting the sequence of SHR values in the formula for CH (6).

For example, CH (1:3 3:1 3:2) with $SHR_{max} = 3$ is converted to RGB (0, 200, 150).

*2.3. Oscillator Circuit*

A model diagram of a single oscillator consists of a current source $I_P$, a capacitance $C$ connected in parallel with the VO$_2$ switch, and a noise source $U_n$ (Figure 7a). The capacitance $C$ remains constant $C = 10$nF, while $I_P$ and $U_n$ vary in the following ranges $I_P$ (435 µA – 1220 µA), $U_n$ (0 mV – 10 mV). The noise source simulates external or internal circuit noise, for example, switch current noise manifested in fluctuations of switch threshold voltages [32]. $I_{sw}$ and $U$ denote the current passing through the VO$_2$ switch and the voltage on it, respectively. The model current-voltage characteristic of the VO$_2$ switch is shown in Figure 7b. All model switches without coupling have the same I-V characteristic, with stationary natural parameters $U_{th} = 5$V, $I_{th} = 550$ µA (threshold voltage and current), $U_h = 1.5$ V, $I_h = 1039$ µA (holder voltage and current), $U_{cf} = 0.82$ V (cutoff voltage). Model curve $I_{sw} = f(U)$ has high-resistance (OFF) and low-resistance (ON) segments with corresponding dynamic resistances $R_{off} = 9100$ Ω and $R_{on} = 615$ Ω. A more detailed description of the model circuit of a coupled oscillator-based neural network and methods for calculating oscillations in a circuit are given in a previous study [26]. The system of differential equations for the current in the circuit were numerically calculated with respect to time at regular intervals $\Delta t = 1$ µs.

Tangibly, the operation of the circuit can be enacted as periodic charging and discharging of the capacitor, when the operating point of the circuit is kept in the negative differential resistance (NDR) section, at the expense of the current source $I_P$.

Examples of the calculated oscillogram sections for voltage $U$ and current $I_{sw}$ for the reference oscillator are shown in Figure 8a.

The dependence of the own oscillations frequency $F_0$ on the magnitude of the current $I_P$ is shown in Figure 8b. Own oscillations occur when the circuit operation point lies in the region of negative differential resistance, in the current range of $I_{th} \leq I_P \leq I_h$ (550 µA $\leq I_P \leq$ 1105 µA), with limit frequencies of 1640 Hz and 10130 Hz. The maximum frequency of 12850 Hz corresponds to a current $I_P = 1039$ µA. In this way, an increase in the supply current $I_P$ initially leads to an increase in frequency, due to a decrease in the charging time of the capacitor $C$ to the threshold turn-on voltage $U_{th}$, then the frequency decreases due to an increase in the discharge time of the capacitor to the holder voltage $U_h$.

By setting the supply current below $I_P < I_{th}$ or above $I_P > I_h$, the operating point of the circuit is set to the sub-threshold state, where own generation is absent and the switch is always either off or on, respectively (see Figure 8b). Further, we will call such an oscillator operation mode a sub-threshold oscillator operation mode.

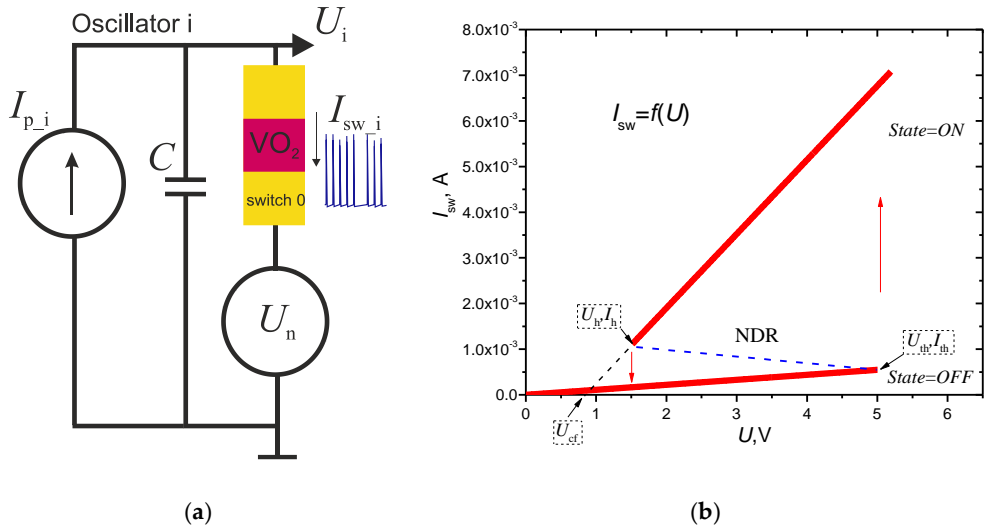

(**a**) (**b**)

**Figure 7.** (**a**) Diagram of a single oscillator based on a VO$_2$ structure. Note: *I* = number of the oscillator; $I_P$ = current source; *C* = capacitance; $U_n$ = noise source; $I_{sw}$ = current passing through the VO$_2$ switch; *U* = voltage on the switch. (**b**) Model I-V characteristic of a separate switch.

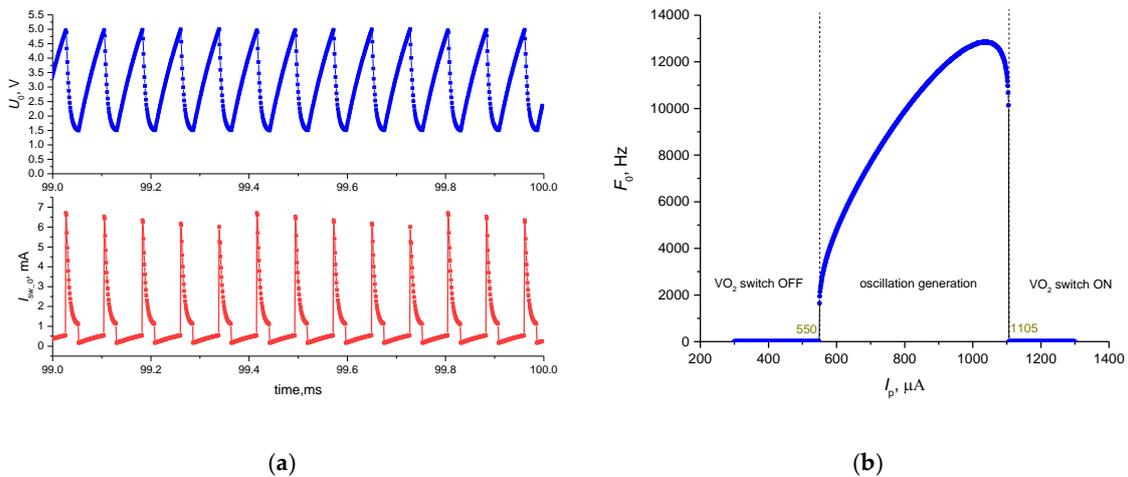

(**a**) (**b**)

**Figure 8.** (**a**) Calculated oscillograms sections for the oscillator 0 at $I_{P\_0}$ = 1039 µA, voltage $U_0$, and current $I_{sw\_0}$. (**b**) Dependence of the single oscillator's own oscillation frequency on the supply current.

### 2.4. ONN Structure

The studied ONN consists of a reference oscillator (No.0), whose frequency does not change, the input layer in the form of a one-dimensional matrix, each element of which is represented by one VO$_2$ oscillator (No.1–4), and the output oscillator (No.5) (Figure 9). Data in the form of binary four-digit numbers is transmitted to a layer of input oscillators in such a way that each oscillator of this layer is associated with two values of supply currents: $I_P$ = $I_{OFF}$ represents logical 0, or $I_P$ = $I_{ON}$ represents logical 1. The numbers can be associated with two-dimensional images, as shown in Figure 9. At the system output, the synchronization order of the output oscillator relative to the reference oscillator is expressed either by the high order synchronization SHR$_{0,5}$ or by the chimeric synchronization index CH$_{0,5}$ (see formula 6). Thus, four-digit numbers are converted to the synchronization state of the output oscillator.

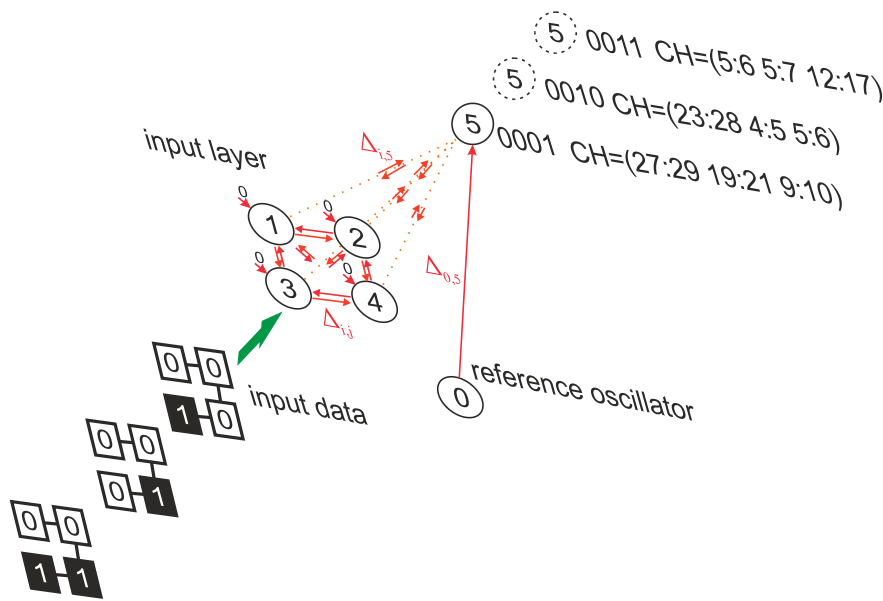

**Figure 9.** Model scheme of ONN with an example of converting binary numbers to CH values of the output oscillator's synchronous state.

The thermal effects of oscillators No.1–5 on each other in the circuit in Figure 9 are set to be mutually equivalent, that is, $\Delta_{i,j} = \Delta_{j,i}$, while the reverse effect on the reference oscillator (No.0) from other oscillators are not taken into account: $\Delta_{1,i} \neq 0$, $\Delta_{i,1} = 0$. Thus, the reference oscillator generates the thermal pulses and sets the rhythm of the entire ONN with a constant frequency.

The switch parameters are unchanged in numerical simulation, while current intensities $I_{p\_i}$ ($I_{ON}$, $I_{OFF}$, $I_{p\_0}$, $I_{p\_5}$), coupling strength constants $\Delta_{j,i}$, noise amplitude $U_n$, and $\eta_{th}$ vary.

*2.5. Task Setting And Technique of ONN Training*

One of the tasks that can be set for the network is to perform conversion functions. When 16 variants of input signals are applied, the output oscillator can take different synchronization values, including showing no synchronization if $\eta < \eta_{th}$, as in Equation (5). Denoting the total number of synchronous states at the output as $x$, $x$ can take a value from 1 to 16. Among the $x$ synchronous states, there can be $n$ unique synchronization values. Then, the output can be codified as "$n$ of $x$". In the case where all of the output synchronization values are unique ($n = x$), this corresponds to the result "$x$ of $x$".

We have divided the network responses into 16 variants, corresponding to the condition when a certain number of synchronous states occur and none of them repeat. For example, the task "1 of 1" corresponds to one synchronous state at the output, all other states are not synchronous, and "2 of 2" corresponds to two different states (see explanatory Table 1).

The most difficult task is the case of "16 of 16", when each input state corresponds to its unique synchronous state, expressed by one of two indices, $SHR_{0,5}$ or $CH_{0,5}$ (see Equation (6)).

The possible number of representations of the task "$x$ of $x$" is equal to the number of combinations x out of 16 ($C_{16}^x$). Therefore, the highest probability to find a solution is for x = 8, and the smallest probability is for x = 16 at $C_{16}^{16} = 1$ (see Figure 10). However, as it will be shown in the results, the type of distribution of the found solutions differs significantly from the distribution in Figure 10 and is determined by the network parameters. The total number of options for presenting tasks "$x$ of $x$" is $\sum C_{16}^x + 1 = 2^{16} = 65536$, where the absence of synchronization with any input data is taken into account. The total number of network response options is much higher if we take into account answers of the type "$n$ of $x$".

**Table 1.** Example of network responses for the "*x* of *x*" task. Empty cells correspond to the absence of synchronization by criterion of Equation (5). The responses are taken at high order SHR$_{0,5}$ and chimeric CH$_{0,5}$ synchronization.

| | High order synchronization | | | Chimeric synchronization | | |
|---|---|---|---|---|---|---|
| | "1 of 1" | "2 of 2" | "16 of 16" | "1 of 1" | "2 of 2" | "16 of 16" |
| Input data | SHR$_{0,5}$ | SHR$_{0,5}$ | SHR$_{0,5}$ | CH$_{0,5}$ | CH$_{0,5}$ | CH$_{0,5}$ |
| 0000 | - | - | 1:1 | (3:4) | - | (7:8 11:12 3:3) |
| 0001 | - | - | 9:7 | - | - | (27:29 19:21 9:10) |
| 0010 | - | - | 9:8 | - | - | (23:28 4:5 5:6) |
| 0011 | - | - | 10:7 | - | - | (5:6 5:7 12:17) |
| 0100 | - | - | 14:11 | - | - | (26:28 25:27 13:15) |
| 0101 | - | 1:1 | 3:2 | - | (13:11 6:5) | (46:48 25:27 6:7) |
| 0110 | - | - | 5:3 | - | - | (4:5 3:4 5:6) |
| 0111 | - | - | 8:5 | - | (19:17 29:26) | (8:10 5:6 9:11) |
| 1000 | - | - | 20:17 | - | - | (5:6 7:8 1:1) |
| 1001 | - | - | 2:1 | - | - | (42:43 39:40 17:19) |
| 1010 | - | - | 11:7 | - | - | (5:7 12:17 9:11) |
| 1011 | - | - | 15:7 | - | - | (5:7 12:17 10:12) |
| 1100 | - | - | 15:8 | - | - | (9:10 13:14 1:1) |
| 1101 | - | 7:8 | 11:5 | - | - | (9:10 6:7 3:3) |
| 1110 | - | - | 19:9 | - | - | (11:15 8:11 5:7) |
| 1111 | 4:3 | - | 5:2 | - | - | (8:11 5:7) |

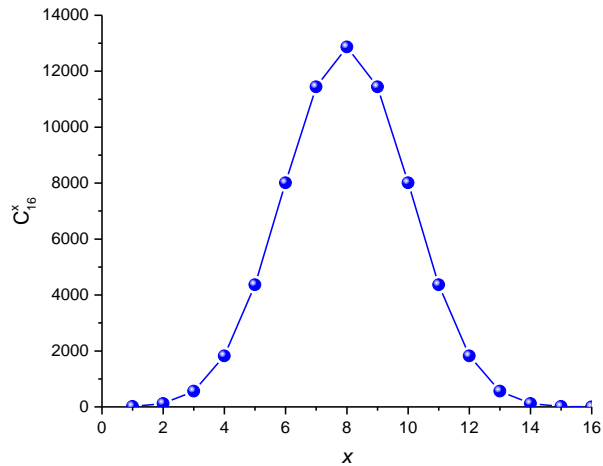

**Figure 10.** Histogram of the number of "*x* of *x*" task representations.

As ONN with the high order synchronization effect has not been studied before, there are no established methods for network training. One of the ways is to use the Simulated Annealing Algorithm (SAA) [1] for the network parameter selection: currents ($I_{ON}$, $I_{OFF}$, $I_{P\_0}$, $I_{P\_5}$), coupling strength $\Delta_{j,i}$, noise amplitude $U_n$, and synchronization effectiveness threshold $\eta_{th}$. The algorithm's key point is the random searching of problem solutions at some initial interval of parameters followed by narrowing of these intervals. In the majority of cases considered in the article, we used only a random search in a given range of parameters. In some cases, a better solution could be found by gradient descent near the found solution. The random search algorithm was tested in a previous study [26] and showed its effectiveness, since a feature of the neural network is the presence of a set of solutions (a set of system parameters) to satisfy the answer for the task "*x* of *x*". Therefore, in the results, we present a distribution histogram of the number of solutions, called NS, from the value x.

The values of $I_{ON}$ and $I_{OFF}$ currents set the logic levels 1 and 0 by determining the supply current of the corresponding input oscillator. The range of current variation is (435 µA – 1220 µA). This range is wider than the range of existence of own oscillations (550 µA – 1105 µA) defined by Figure 8b in Section 2.1. If we represent a pair of currents ($I_{ON}$ and $I_{OFF}$) in the diagram by the point, as in Figure 11, then the central area 1 corresponds to the mode where both logic levels make the oscillators oscillate. In other words, when applying either 1 or 0, the oscillator is in the generation zone, and regardless of the input signal, all oscillators oscillate. In area 3, currents $I_{ON}$ and $I_{OFF}$ lead the oscillator into sub-threshold mode, while the switch is either ON or OFF.

Area 2 corresponds to the case when one of the current levels $I_{ON}$ or $I_{OFF}$ leads to generation, while the other level sets the oscillator to the sub-threshold mode.

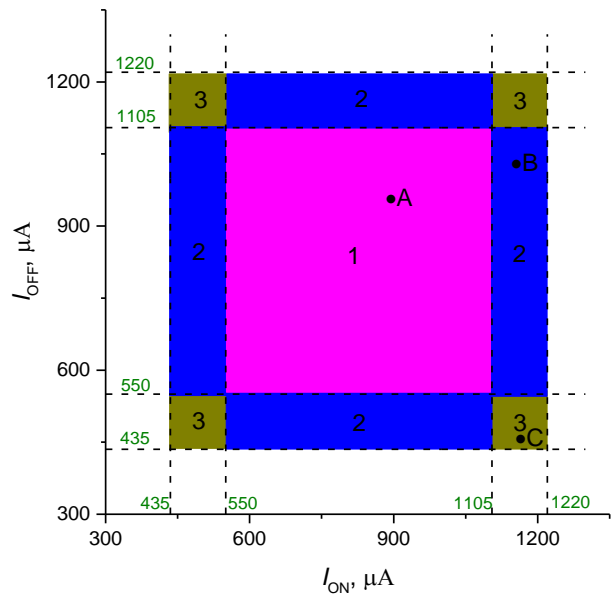

**Figure 11.** Ranges of current variation for $I_{ON}$ and $I_{OFF}$ with division into areas (area 1, area 2, area 3).

For example, point **A** in Figure 11 corresponds to currents $I_{ON}$ = 894 µA and $I_{OFF}$ = 958 µA, and both levels of current lead to oscillation. At point **B** ($I_{ON}$ = 1153 µA and $I_{OFF}$ = 1036 µA), a logical 1 ($I_{ON}$) leads the oscillator to enter the threshold mode, while the switch is in the on mode, its resistance is $R_{ON}$, and a large current passes through it. In turn, a logical 0 ($I_{OFF}$) leads to the generation of oscillations. Point **C** ($I_{ON}$ = 1163 µA and $I_{OFF}$ = 459 µA) corresponds to the mode when both 1 and 0 set the oscillator to the sub-threshold mode, while at 1 the switch is on, and at 0 the switch is off.

In the sub-threshold mode, the switch is turned on and heats the neighboring areas. Therefore, the switch affects the threshold voltages of the neighboring areas and plays the role of a constant displacement neuron. In the model, this is inherently taken into account, as the influence of oscillators is affected by the state of the switches (see Equations (A1)–(A3) in the previous study [26]).

The range of variable currents is chosen in a way that area 1 (pink space) is equal to the sum of areas 2 (blue) and 3 (brown). Therefore, with a random choice of a point in the space of the currents $I_{OFF}$ and $I_{ON}$, the probabilities of falling into the region of oscillation generation and into the region of the sub-threshold state of the oscillators are equal, which allows us to compare histograms of the solutions NS.

The ONN was set up by brute force, searching the thermal coupling strength $\Delta_{i,j}$ and power supply levels ($I_{OFF}$ and $I_{ON}$) of the input layer oscillators (No.1–4). The random search was performed in the ranges of 0–1 V for $\Delta_{i,j}$ (except $\Delta_{0,j}$ = 0.2 V) and 435 µA – 1220 µA for $I_{OFF}$ and $I_{ON}$. The values of the supply currents of the reference $I_{p\_0}$ and output $I_{p\_5}$ oscillators, and the noise level $U_n$ were fixed. The number of samples was $10^5$.

In the current study, the target was to find solutions "*x* of *x*" defining the operation of the neural network converter, as well as the use of this scheme for filtering images (see Section 3.6). Other tasks may require different network functions. For example, in the task of driving a vehicle, proximity sensors can input the signals of approaching an obstacle, and output synchronization would determine in which direction the vehicle should turn.

*2.6. Method of Oscillators Coupling and Variants of Experimental Implementation*

The main model elements of the oscillator network are VO$_2$ switches, which are two-electrode planar structures with a functional layer of vanadium dioxide and two metal contacts. The interaction between the oscillators is carried out by heat flows propagating through the substrate, resulting from the Joule heating when the switches are turned on. This method of coupling was experimentally demonstrated in a previous study [24].

The interaction of oscillators through thermal coupling assumes a reduction in the threshold switching voltage of each switch $U_{th}$ by a coupling strength value $\Delta_{i,j}$ when the switches thermally affect each other at the moment of capacitance $C$ discharge and the release of Joule heat. A detailed mathematical model of thermal coupling of VO$_2$ oscillators is given in a previous study [26].

An exemplary view of the location of the switches on the substrate is shown in Figure 12. In the study, we modeled the reference oscillator, which affected all other oscillators, and unidirectional thermal couplings. The question may arise of how to implement unidirectional thermal coupling in a real experiment. The task of the technical implementation of the proposed model objects is the subject of a separate study; nevertheless, possible variants of one-way thermal coupling between two oscillators are presented in Figure 13. In the diagram (Figure 13a), the connection is carried out through an additional current resistor included in the circuit and located on the substrate along with switches. Its resistance does not depend on temperature, however, when heated by the current, it transmits a thermal effect from oscillator 0 to oscillator 1. Therefore, the implementation of one-way thermal coupling is possible, in principle. In addition, the amplitude of the thermal interaction signal decays exponentially with the distance between the switches; see detailed descriptions of the physics of thermal coupling in previous studies [24,33]. Therefore, the one-way connection can be accomplished by making switches with different radii of thermal exposure $R_{TC}$, as in Figure 13b.

The thermal coupling is chosen due to simplicity of modelling in the case of a fully connected circuit of oscillators; oscillators are galvanically isolated, and this type of coupling has been confirmed experimentally.

In addition to thermal coupling, oscillators can be connected electrically via resistors or capacitors [34], optically [35,36], and via wireless channels [37], however, the method for determining chimeric synchronization is universal, regardless of the coupling type and the physics of the oscillators.

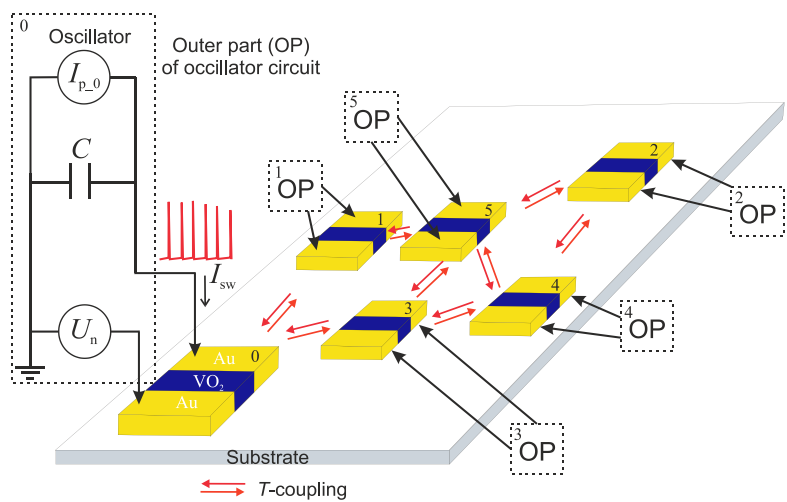

**Figure 12.** Electrical circuit of thermally coupled VO$_2$-oscillators located on the substrate. The numbers of the VO$_2$ structures correspond to the numbers of the oscillators. Double arrows indicate a thermal bond.

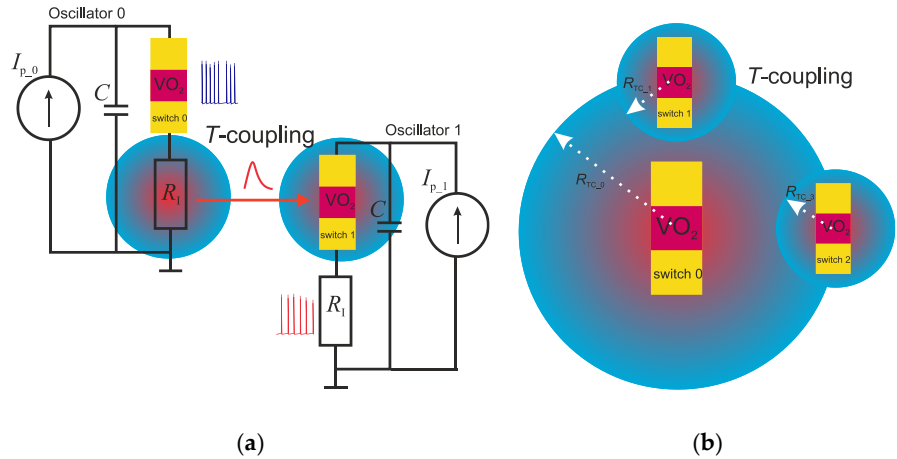

**Figure 13.** Variants of the technical implementation of one-way thermal coupling: (**a**) through an additional current resistor, (**b**) by making switches with different radii of thermal exposure $R_{TC}$.

## 3. Results

### 3.1. Investigation of the Chimeric Synchronization of Two Coupled Oscillators

Let us consider how the distribution of synchronous states of two oscillators changes in the space of supply currents with the parameters $\eta_{th}$ = 90%, $U_n$ = 1 mV, $\Delta_{0,5}$ =0.2 V, $\Delta_{5,0}$ =0 V (Figure 14a). Oscillators are connected by one-way coupling; oscillator No.0 affects oscillator No.5, and oscillators No.1–4 in the diagram in Figure 14a are disabled. Figure 14b shows the synchronization distribution in the form of Arnold tongues, when the chimeric synchronization is not taken into account (the colors correspond to the procedure described in Section 2.2).

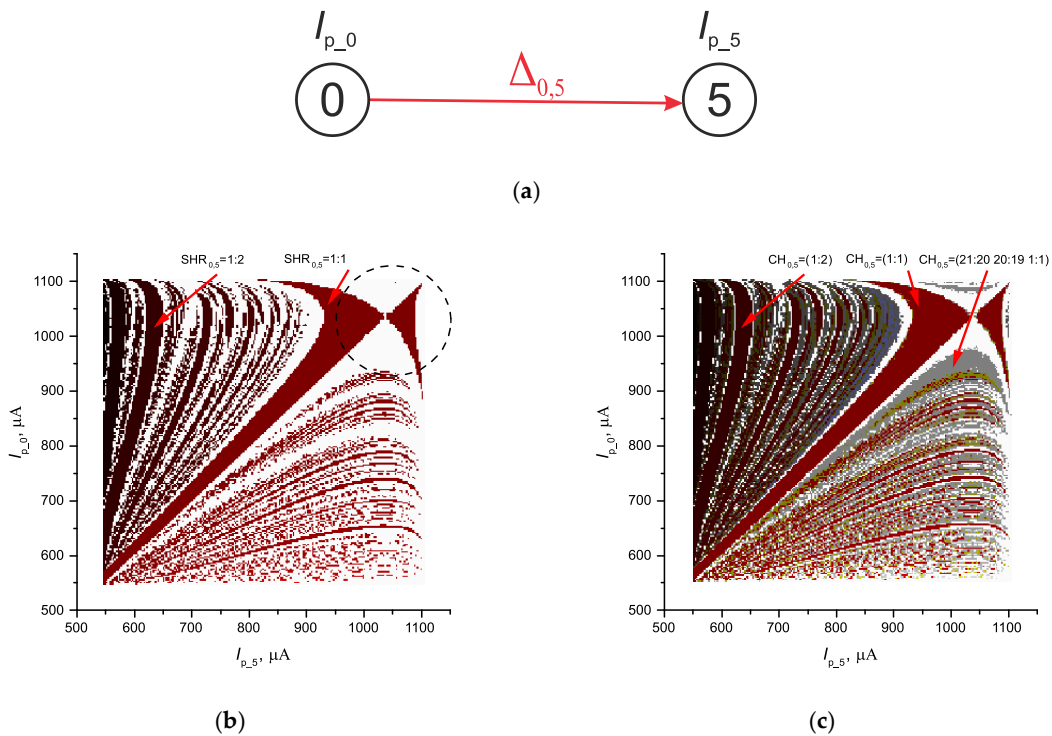

**Figure 14.** Schematic mapping of two coupled oscillators with one-way coupling (**a**), distribution of their synchronization in the space of supply currents without considering chimeric synchronization (**b**), and considering the chimeric synchronization (**c**). Colors are calculated by the method presented in Section 2.2, where SHR$_{max}$ = 4. The dotted line indicates the nonlinear frequency domain of the oscillator, and the arrows indicate the sample values of $SHR_{0,5}$ and $CH_{0,5}$. The values of the parameters are $\eta_{th}$ = 90%, $U_n$ = 1 mV, $\Delta_{0,5}$ = 0.2 V.

Arnold tongues have a linear extended form, however, they are transformed at high currents and shape regions with alternating synchronization (indicated by a dotted line). |This behavior is explained by the nonlinear dependence of the frequency on the supply current, as shown in Figure 8b. In addition, a well-known phenomenon is observed when synchronization occurs predominantly in areas where the frequency of oscillator No.0 is greater than the frequency of oscillator No.5. This is clearly seen for synchronization $SHR_{0,5}$ = 1:1, which occurs predominantly above the diagonal, as seen in Figure 14b. The unidirectional action of oscillator 0 is more effective in this case, as it leads and initiates oscillations in the adjacent circuit.

Application of the technique described in Section 2.1 makes areas with chimeric synchronization visible, indicated by the gray color in Figure 14c. For example, the state $CH_{0,5}$ = (21:20 20:19 1:1) occurs at the currents $I_{P\_0}$ = 948 μA, $I_{P\_5}$ =1004 μA, and the synchronization efficiency $\eta$ = 98%. In general, the areas with detected synchronization with the efficiency $\eta \geq \eta_{th}$ are increased, due to the appearance of areas with chimeric synchronization. Now, the areas previously diagnosed as nonsynchronous operating modes of oscillators have a well-defined classification with chimeric synchronization, and can be used in creation of logical and neuromorphic devices.

*3.2. The Study of the Neural Network Information Converter, without Accounting for Chimeric Synchronization*

The ONN under investigation has the structure described in Section 2.5. The output oscillator's state is determined by the conventional synchronization index $SHR_{0,5}$. The training was carried out according to the method described in Section 2.6, with the number of samples set as $10^5$.

The results of the distribution of the number of NS solutions for the "*x* of *x*" problem after training are shown in Figure 15a. Four curves are presented; the curve ("whole area", black marker) corresponds to the entire space of the $I_{ON}$-$I_{OFF}$ currents, and the remaining curves correspond to different areas of the $I_{ON}$-$I_{OFF}$ space (see Figure 11). The calculation was performed at a noise level of $U_n$ = 0 mV, $\eta_{th}$ = 30% and $I_{P\_0}$ = 1039 μA, $I_{P\_5}$ = 750 μA.

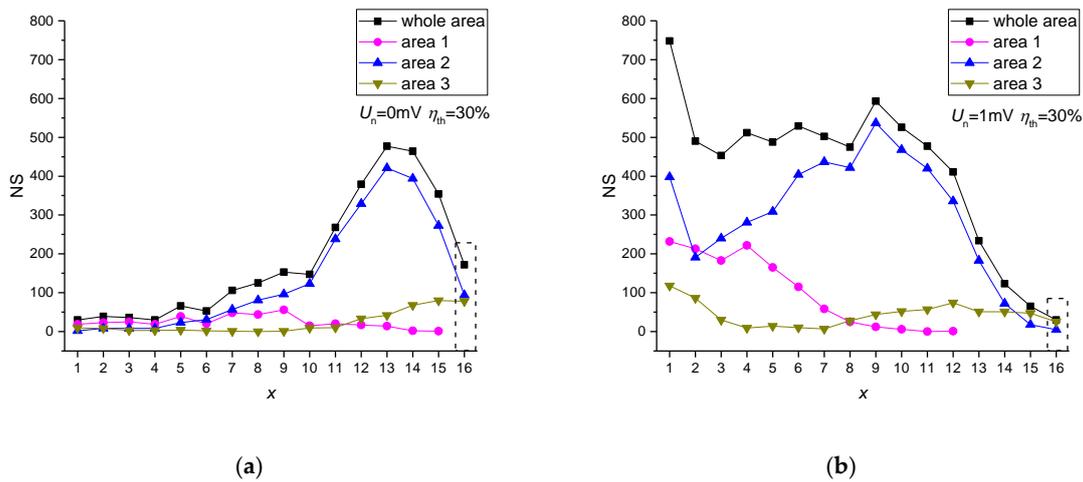

**Figure 15.** Distribution of the number of NS solutions to the "*x* of *x*" problem, for different regions of the $I_{ON}$-$I_{OFF}$ current space without accounting for chimeric synchronization at $U_n$ = 0 V (**a**) and $U_n$ = 1 mV (**b**). The numbers of the regions correspond to Figure 11. The dotted line highlights the area of the solutions to the "16 of 16" problem.

It can be seen that the solution to the "16 of 16" problem is achieved only in regions 2 and 3 (see the section marked by the dotted line $NS^{2,3}(16) \sim 100$), when one of the oscillators is in the sub-threshold mode. No solution $NS^1(16)$ = 0 is found in area 1. The maximum number of solutions belongs to the "13 of 13" task, corresponding to area 2 ($NS^2(13) \sim 400$). For area 1, the maximum number of solutions falls within the task "9 of 9" ($NS^1(9) \sim 50$). In area 2, there are more solutions than in all other areas, which means the circuit has the greatest potential when one of the currents ($I_{ON}$ or $I_{OFF}$) sets the oscillator into the threshold state, and the other current creates the generation mode of the oscillator.

The type of distribution depends on a number of parameters, such as $U_n$, $\eta_{th}$, $I_{P\_0}$, $I_{P\_5}$. When adding noise to the system $U_n = 1$ mV (Figure 15b), the number of solutions to the "16 of 16" problem significantly decreases ($NS^{2,3}(16) \sim 10$), but the number of solutions with a low $x$ significantly increases. Therefore, noise can increase the number of solutions, which is associated with the decrease of solutions in one range and an increase in another. Apparently, there is an optimal noise value when the maximum number of solutions is observed (since with an unlimited increase in noise, the number of solutions will obviously decrease), and this effect is similar to the stochastic resonance effect observed in a previous study [26].

*3.3. The Study of the Neural Network Information Converter, Accounting for Chimeric Synchronization*

The distribution results of the number of NS solutions to the "$x$ of $x$" problem after training are presented in Figure 16a. Inclusion of chimeric synchronization leads to an increase in the NS maximum by more than 10 times, and the number of solutions to the "16 of 16" problem increases by 30 times. The greatest number of solutions, as in the previous case, falls on area 2. Another interesting result is the appearance of solutions in area 1, where all generators are active at any input data.

Hence, accounting for chimeric synchronization significantly increases the probability of finding solutions.

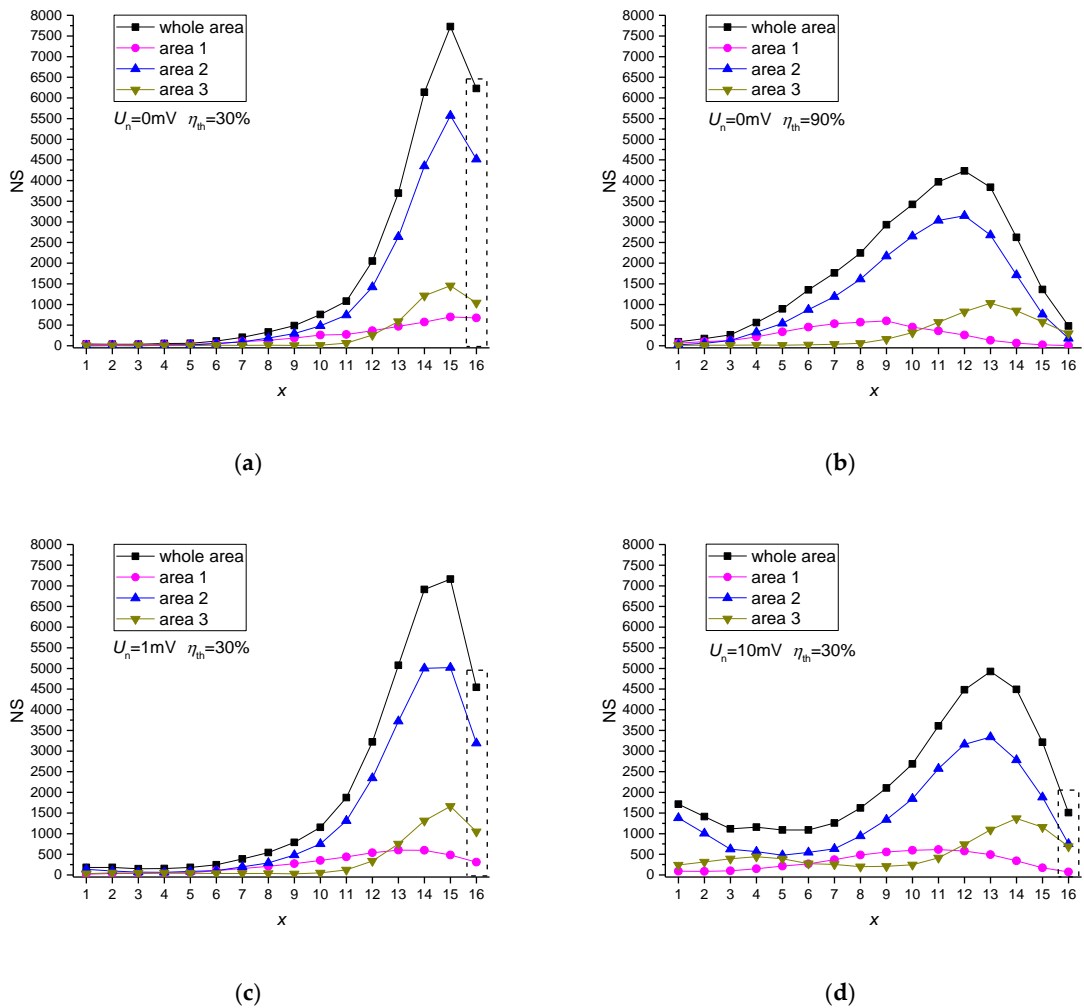

(a)    (b)    (c)    (d)

**Figure 16.** Distribution of the number of NS solutions to the "*x* of *x*" problem for different areas of the $I_{ON}$-$I_{OFF}$ space, taking into account chimeric synchronization. Calculation parameters are $I_{P\_0}$ = 1039 µA, $I_{P\_5}$ = 750 µA, (**a**) $U_n$ = 0 mV, $\eta_{th}$ = 30%, (**b**),# $U_n$ = 0 mV, $\eta_{th}$ = 90%, (**c**) $U_n$ = 1 mV, $\eta_{th}$ = 30%, (**d**) $U_n$ = 10 mV, $\eta_{th}$ = 30%. The numbers of the regions correspond to Figure 11. The dotted line highlights the area of the solutions to the "16 of 16" problem.

Four curves are presented; the curve (black marker) corresponds to the entire space of the $I_{ON}$-$I_{OFF}$ currents, and the other curves correspond to different areas of the $I_{ON}$-$I_{OFF}$ space (see Figure 11). The distribution of solutions to the "16 of 16" problem in the $I_{ON}$-$I_{OFF}$ space is presented in Figure 17a. The distribution is symmetrical to the diagonal and concentrated in area 2. No solutions exist when both currents are greater than 1105 mA or less than 550 mA. In this case, both the $I_{ON}$ and $I_{OFF}$ currents lead to either switching on or switching off the switch and set the oscillator to the subthreshold mode, and there is no dependence of current influence on the network from the input data.

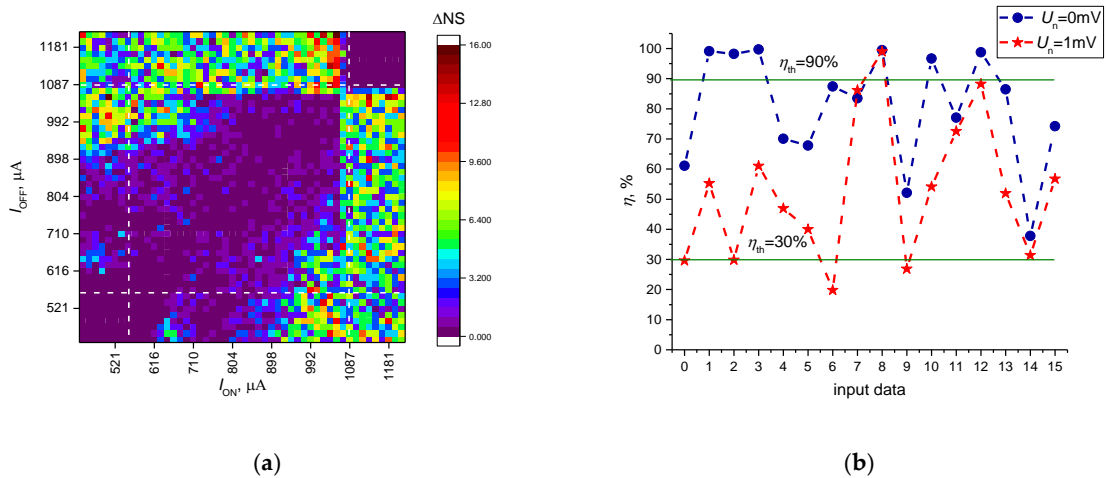

(**a**) (**b**)

**Figure 17.** (**a**) The density of distribution of solutions to the "16 of 16" task in the $I_{ON}$ and $I_{OFF}$ current space, dotted lines indicate the boundaries of the areas (area 1, area 2, area 3 in accordance with Figure 11). (**b**) The dependence of the synchronization efficiency $\eta$ on the input data (in decimal format) at two amplitudes of internal noise $U_n$.

The synchronization efficiency $\eta_{th}$ and the noise level $U_n$ affect the form of the distribution NS. Figure 16b presents the NS distribution for $\eta_{th}$ = 90%. The distribution maximum is shifted to the left, and the maximum of the total number of solutions is in the "12 of 12" area. At the same time, the maximum value decreased by approximately 2 times, and the number of solutions to the "16 of 16" task fell by 10 times. Such a decrease happens due to the cut-off of solutions by the criterion $\eta \geq \eta_{th}$. In Figure 17b (blue dots), the dependencies of the synchronization efficiency values on the input data are presented, which are elements of a single solution. So, at $\eta_{th}$ = 30% ($U_n$ = 0 mV) all points satisfy the condition $\eta \geq \eta_{th}$, and the array of elements $CH_{0,5}$ satisfies the "16 of 16" problem. At the same time, at $\eta_{th}$ = 90% only six points satisfy the $\eta \geq \eta_{th}$ condition and constitute a solution for the "6 of 6" task. Therefore, with the increase of $\eta_{th}$, the number of solutions to "*x* of *x*" problems with lower *x* grows and shifts the NS distribution to the left, while NS decreases for the "16 of 16" task.

Figure 16c,d represents the NS distributions for different noise levels $U_n$.

An increase in noise leads to a decrease in the number of solutions to the "16 of 16" problem, with the distribution shifted to the left. In contrast to the influence of $\eta_{th}$, an increase in noise leads to a significant increase in the number of solutions to "*x* of *x*" problems with lower *x*, and NS experiences the most significant increase for the "1 of 1" task. The noise has the least impact on solutions in area 3, as the input layer oscillators here are constantly in the sub-threshold stable state regardless of the noise voltage dynamics, and the noise affects only the reference oscillator No.0 and output oscillator No.5. With increasing noise (Figure 17b), the synchronization efficiency of almost all elements of the solution decreases, as reflected in the observed patterns of NS distribution.

*3.4. Scheme of Converting with a Minimum Number of Couplings*

A special case of a fully connected circuit is the circuit shown in Figure 18a, which contains only unidirectional links with the output oscillator, and the input oscillators do not interact.

The scheme is simple and contains only five couplings, however, its functionality allows solving the "*x* of *x*" tasks. The distribution of the number of NS solutions to the "*x* of *x*" problem after training is shown in Figure 18b. Solutions exist in all areas, however, the number of solutions is less by 2 times than for a fully connected network (see Figure 16a). The maximum of solutions at $\eta_{th}$ = 30% and $U_n$ = 0mV falls on the "16 of 16" task.

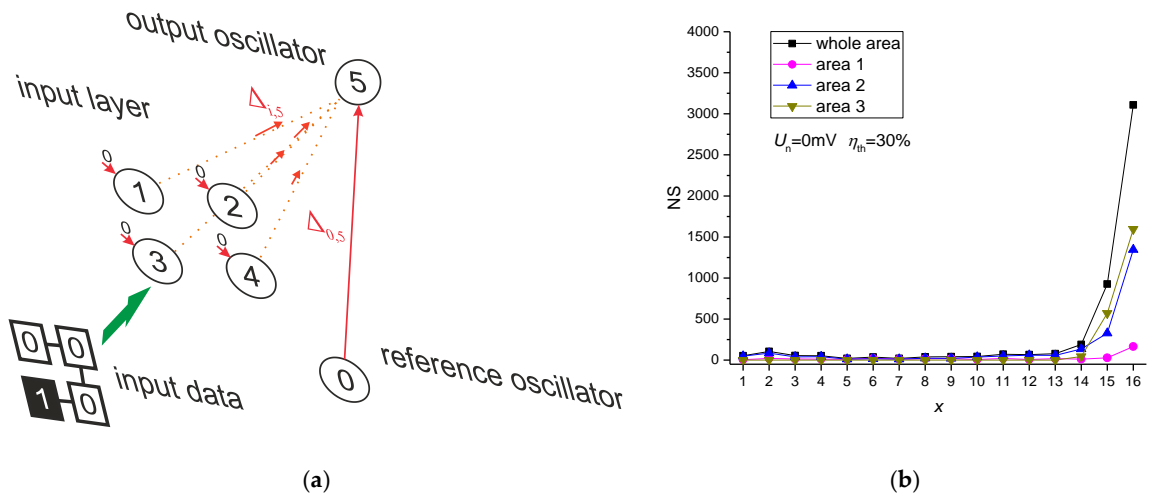

(**a**) (**b**)

**Figure 18.** (**a**) Model scheme of ONN containing only unidirectional couplings of oscillator No.0-4 with output oscillator No.5, and (**b**) distribution of the number of NS solutions to the "*x* of *x*" problem.

Another important conclusion is that the input signal can be represented not by the levels of the $I_{ON}$-$I_{OFF}$ currents, but by signals with a certain $CH_{0,i}$. In this way, the input and output of the network will be signals of the same nature—signals with a certain synchronization (see Figure 19c). This feature is important for designing large-scale networks, where the output of one network can become the input for another network, as in transistor–transistor logic (TTL) circuits. Next, we summarize the concept of devices that can be implemented on the studied scheme.

*3.5. Technological Concepts of Neural Network Converters*

This section provides options for the technological concepts of neural network converters. Figure 19a shows a diagram based on 6 oscillators. In this scheme, current levels are applied to the input, and the output is a signal with a certain chimeric synchronization. The circuit requires the generation mode of one of the input oscillators, therefore, the input currents must belong to areas 1 or 2 (see Figure 11).

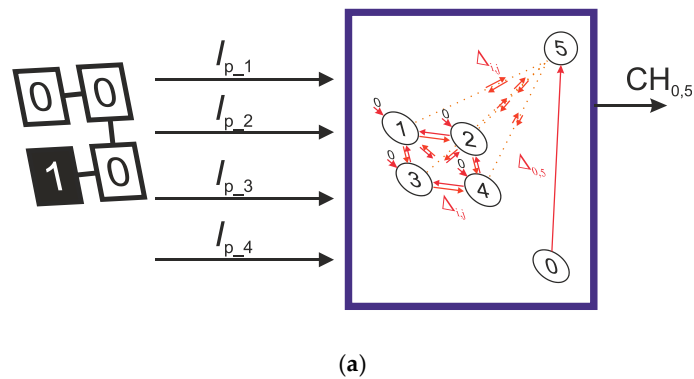

(**a**)

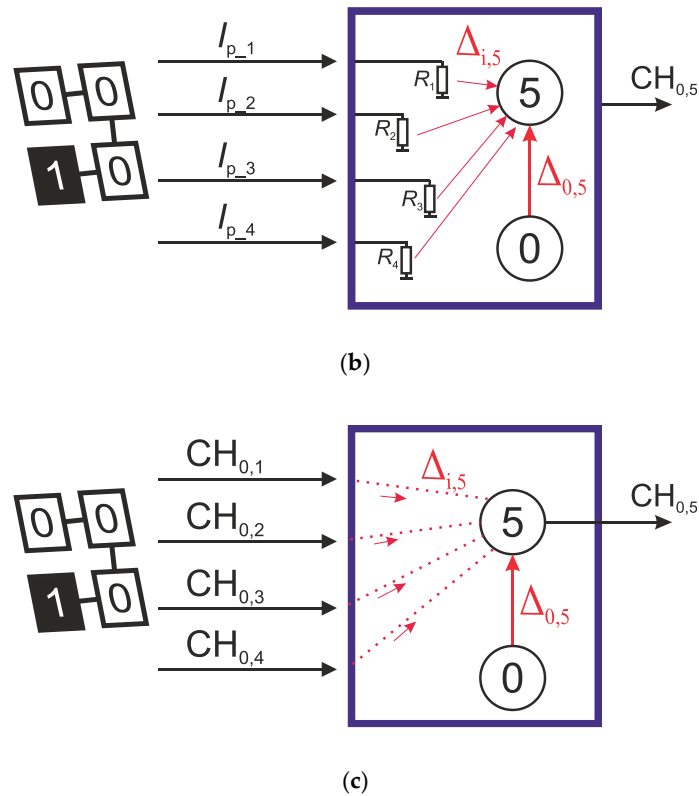

**Figure 19.** (**a**) A fully connected network, where the input signal is determined by oscillator currents $I_P$ and the output signal is determined by synchronization value CH. (**b**) A simplified two-oscillator circuit, where the input signal is given by the currents of thermistors. (**c**) A simplified two-oscillator circuit, where the input and output signals consist of signals of a certain synchronization.

The scheme in Figure 19b consists of only two oscillators, and the input signal is given by the currents of the thermistors. In fact, this is an imitation of the operation of the circuit on Figure 19a in area 3 of the current parameters. Thermistors set the offset of the threshold characteristics of oscillator No.5 and in this way change the output synchronization. The advantages of the scheme are a lower number of oscillators and increased resistance to noise.

The scheme in Figure 19c differs from the previous schemes, and the principle of its operation is analyzed in Section 3.4. At the input and output of the network there are signals of the same nature—signals with a certain synchronization. This is important for designing large-scale networks, where the output of one network becomes the input for another network, similar to TTL logic circuits.

The coupling forces $\Delta_{i,j}$ determine the input weights and largely determine the functionality of all devices.

*3.6. Examples of Image Processing*

The converters described in the previous paragraph can be used to filter images. As shown in Figure 19, the input data can be a 2 × 2 matrix of pixels. For example, fully filled pixels correspond to the input number "1111", and the absence of any filled pixels constitutes the number "0000". A defined pixel color can be associated with the output. By applying a 2 × 2 matrix transformation algorithm sequentially to all image pixels, it is possible to filter certain attributes. Table 2 demonstrates the variants of the simplest filters that allow removal of noise from an image (Filter 1) and a filter for searching the borders of objects (Filter 2). Filter 1 is a special case of the "1 of 1" task, and Filter 2 is the "2 of 2" task.

**Table 2.** Examples of noise filter (Filter 1) and object boundaries filter (Filter 2).

|            | Filter 1 | Filter 2 |            | Filter 1 | Filter 2 |
|------------|----------|----------|------------|----------|----------|
| Input Data | Color    | Color    | Input Data | Color    | Color    |
| 0000       | White    | White    | 1000       | White    | Black    |
| 0001       | White    | Black    | 1001       | White    | Black    |
| 0010       | White    | Black    | 1010       | White    | Black    |
| 0011       | White    | Black    | 1011       | White    | Black    |
| 0100       | White    | Black    | 1100       | White    | Black    |
| 0101       | White    | Black    | 1101       | White    | Black    |
| 0110       | White    | Black    | 1110       | White    | Black    |
| 0111       | White    | Black    | 1111       | Black    | White    |

For example, the input image is a set of shapes with a noisy background (Figure 20a), then using filter 1, the clear image is shown in Figure 20b. After applying filter 2, the image Figure 20b is converted to the form of defined object boundaries (see Figure 20c).

In this way, logical neuromorphic devices can be successfully used for image and video processing.

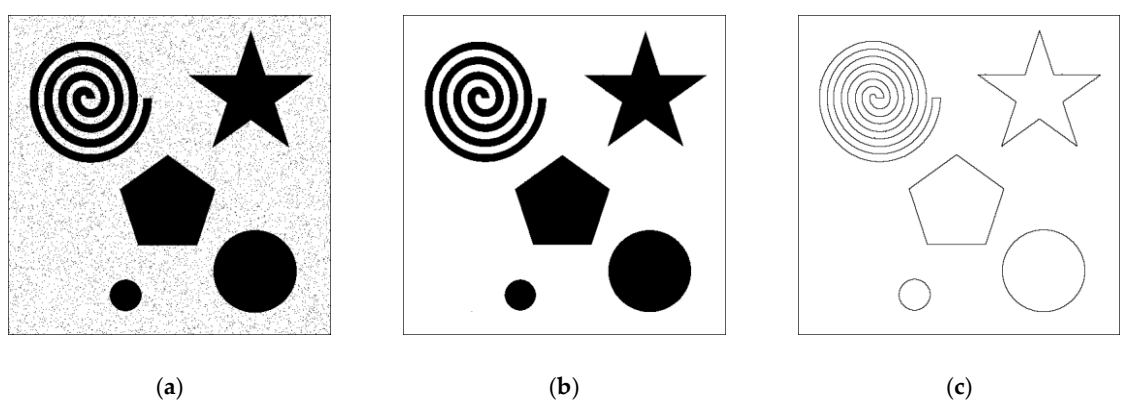

(**a**) (**b**) (**c**)

**Figure 20.** The initial noisy image (**a**), after filtering out noise (**b**), and after finding the boundaries of the figures (**c**).

## 4. Discussion

The random brute force search method used in this pape, is a part of Sample Average Approximation (SAA) method used to optimize neural networks. The method is a statistical algorithm to be applied when the network state is unknown during weighting factors and input parameters changes, or when the neuron activation function is too complex and does not have a specific derivative [1]. For the studied oscillatory neural network, the output parameters are synchronization metrics (CH, SHR, $\eta$), determined by approximately 1000 oscillations; therefore, the functional dependence of the output parameters is difficult to suggest and the application of the gradient approach is limited. The advantage of the SAA method over the Back Propagation Algorithm (BPA), which uses the gradient approach, is the ability of SAA to find the global minimum of the system (global optimization), while BPA tends to get stuck at local minima [38].

Using the brute force search method, the number of satisfying solutions NS for some problems exceeds $10^3$ from $10^5$ random variants (see Figure 16), which is equivalent to the probability of finding a solution of more than 1%. In addition, the use of chimeric synchronization increases the probability of finding solutions compared with the previously described method [26]. This confirms the effectiveness of the brute force search method of the parameters for optimization of the oscillatory neural network.

Further improvements in the network optimization algorithm can be associated with reduced time spent on calculating metrics (CH, SHR, $\eta$). This can be done by simplifying the oscillator model. In addition, the search can be accelerated by using high-performance computing systems with a parallel architecture, for example, GPU and FPGA. The development of optimal and energy-efficient network training algorithms is a task for future research, as the complexity of the oscillator network and the number of its elements will obviously increase in accordance with the complexity of the tasks solved by the network.

The paper demonstrates that the noise value $U_n$ is a controlling parameter, and its value affects synchronization efficiency, which in turn affects the distribution of NS solutions. Figures 15 and 16 illustrate the change in the distribution of NS solutions in the absence of noise in the system at $U_n = 0$ mV. In a real experiment, the noise value, determined by the fluctuation of the threshold turn-on voltage $U_{th}$, can reach tens of millivolts [32,33]. Nevertheless, the described neural network can be used not only as a circuit solution for oscillators based on real switch structures, but also as a model to be implemented in a numerical experiment on computing platforms (GPU, FPGA). In the latter case, the noise level can be model and its value can be varied, starting from $U_n = 0$ mV, depending on the tasks to be solved.

The problem of accounting for the interaction time delay in neural networks [39], and in general, in multi-agent systems [40], has been discussed in the literature. This paper does not take into account the effects of the oscillators' time lag to simplify the understanding of the chimeric synchronization concept. The thermal coupling delay time for the synchronization process of two oscillators is discussed in a previous study [41]. Results have found a time shift between synchronous pulses, and the synchronization efficiency decreases with increasing time delay. The synchronization value SHR remains unchanged, which is consistent with the general findings for multi-agent systems [40] and networks with harmonic oscillators [42]. In addition, the time delay can be ignored if its value is much less than the oscillation period [41].

The stability of chimeric synchronization in the case of the individual oscillator's failure or the presence of interference represents an interesting topic. Several studies investigated the problems of the stable functioning of multi-agent systems [43,44], and in particular, in the field of oscillatory neural networks [45,46], where resistance to variation of natural frequencies, coupling forces, and scaling [46] are in the focus.

The results of this study can be applied to widely known networks based on Kuramoto phase oscillators [47] by using the idea of universal high-order synchronization [47] and chimeric synchronization effects. Unlike pulsed oscillators, where a sharp pulse front due to the switching phenomenon of a nonlinear element can be observed, phase oscillators require an original method for calculating the parameters SHR, CH, and $\eta$. The sharp fronts are easily recorded and processed by technical systems. In the literature, the definition of the order parameter for Kuramoto networks can be found [48], which characterizes the synchronization level and can serve as an analogue of the synchronization efficiency value $\eta$ (4). In addition, the concept of critical communication strength [48–50] correlates with the concept of the threshold synchronization efficiency $\eta_{th}$ (5).

The value of SHR (Equation (1)) determines the pattern of high order synchronization of two signals. The CH equation (Equation (3)) defines a more complex pattern of synchronization and its generalized characteristics; however, CH does not include all the subtleties of the synchronization pattern. The histogram in Figure 6b demonstrates two components of the chimeric synchronization pattern $SHR^1_{i,j}$ = 3:2 and $SHR^2_{i,j}$ = 7:5, cyclically repeating in time as {7:5, 3:2, 3:2}. In a general case, the order patterns in the time scan can be different, depending on the parameters of the system, for example {7:5, 3:2}, {7:5, 3:2, 3:2, 3:2}, or {7:5, 7:5, 7:5, 3:2}. In the latter case, CH can change from $CH_{i,j}$ = (3:2 7:5) to $CH_{i,j}$ = (7:5 3:2) and have a different synchronization efficiency. Therefore, the study of the fine structure patterns of high order chimeric synchronization may be the purpose of future studies.

Chimeric synchronization is the synchronization of quasi-periodic oscillations with a specific synchronization pattern, which determines the intermediate state between full synchronization and the asynchronous state. The transition from a synchronous to asynchronous state is observed when crossing the borders of the Arnold tongue, and a gradual broadening of the spectrum of signals and blurring of phase portraits [18,21] are usually witnessed, as demonstrated in Figure 6. This transition mechanism is called a period doubling bifurcation (or torus-doubling bifurcation) [18,21].

The described method allows the calculation of a family of synchronization metrics between two quasi-periodic sequences of impulses (impulse trains). In other words, the values of SHR and CH can be estimated relative to any two oscillators in the network. In the current study, one of the oscillators is a reference oscillator with a constant frequency to ensure correct functioning of the information converting device.

Chimeric synchronization can be estimated between any number of oscillators. In this case, the representation of the indices (SHR, CH) will be more complicated. For example, for three oscillators in a previous study [25], the SHR index in the form of the ternary fraction $SHR_{i,j,z} = M_i:M_j:M_z$ was used. The study of multi-oscillatory chimeric synchronization could be the subject of future research.

The uniqueness of synchronization metrics (SHR, CH) is in its expression in terms of the ratio of integers (see formulas 1.3), and in the case of the restriction of the integers, the result is discrete. In a previous study [51], when the values of $M_i$, $M_j$ in Equation (1) were restricted in a way that $M_i \leq 20$ and $M_j \leq 20$, the total number of variants $SHR_{i,j}$ was 255. Discreteness of the output parameter is an advantage for technical applications and for interfacing neural networks with computer modules.

An assumption can be proposed that there is a connection between the physics of the formation of chimeric states and the phenomenon of chimeric synchronization. In a previous study [51], a chain of coupled oscillators and long-range synchronization were studied. Results indicate that high order synchronization between a pair of remote oscillators occurs even when intermediate oscillators in connecting links are not formally synchronized. Therefore, it can be assumed that synchronization in these links does exist, but it has a chimeric nature. If, in determining the chimeric states, the chimeric synchronization indices are calculated according to the presented method, the new patterns of the physics of chimeric state formation may emerge.

The method for estimating chimeric synchronization described in this paper gives an idea of the complex structure of synchronization of quasi-periodic oscillations, and can be used to design neuromorphic devices.

## 5. Conclusions

In this study, we have introduced the method for determining chimeric synchronization expressed by a family of metrics. This technique allows reliable determination of the synchronization state and provides an understanding of the complex structure of synchronization of quasi-periodic oscillations, compared to the previously proposed method, which uses a single high-order synchronization parameter [26]. In addition, such a metric can significantly expand the capabilities of neuromorphic and logical devices that operate on the synchronization effect, and increases the number of solutions to the problems posed. We have revealed the distribution of the number of solutions depending on the operating mode of the oscillators, sub-threshold mode, or generation mode. The implementation variants of unidirectional thermal coupling of oscillators are proposed. The thermal mode of oscillators interaction for 3D integration of $VO_2$ switches has an advantage, as thermal waves propagate radially from the switch in all directions along the substrate, and allow thermal coupling of oscillators located on different layers.

Several concepts of the neural network converter have been proposed, in which the input signals are either current or the signals have a certain chimeric synchronization. Examples of the converter application for filtering images have been elaborated.

Finally, the methods for classifying chimeric synchronization and neural network information conversion are universal and can be implemented in ONN with electrical coupling between $VO_2$ oscillators and in ONN based on any switching elements, such as thyristors, tunnel diodes, and resistive memory cells. In addition, these techniques are applicable for implementation on high-performance computing platforms with parallel and variable architecture (CPU, GPU, FPGA), and can be widely used in artificial intelligence systems.

**Funding:** This research was supported by the Russian Science Foundation (grant no. 16-19-00135).

**Acknowledgments:** The author expresses his gratitude to Dr. Andrei Rikkiev for the valuable comments in the course of the article translation and revision.

**Conflicts of Interest:** The authors declare no conflict of interest.

## References

1. Callan, R. *The Essence of Neural Networks*; Prentice Hall Europe: New Jersey, NJ, USA, 1999; ISBN 013908732X.
2. Bishop, C.M. *Neural Networks for Pattern Recognition*; Clarendon Press: New York, NY, USA, 1995; ISBN 0198538642.
3. Roska, T.; Chua, L.O. The CNN universal machine: An analogic array computer. *IEEE Trans. Circuits Syst. II Analog Digit. Signal Process.* **1993**, *40*, 163–173, doi:10.1109/82.222815.
4. Swingler, K. *Applying Neural Networks: A Practical Guide*; Academic Press: California, CA, USA, 1996; ISBN 9780126791709.
5. Kuramoto, Y.; Battogtokh, D. Coexistence of coherence and incoherence in nonlocally coupled phase oscillators. *Nonlinear Phenom. Complex Syst.* **2002**, *5*, 380–385.
6. Abrams, D.M.; Strogatz, S.H. Chimera states for coupled oscillators. *Phys. Rev. Lett.* **2004**, *93*, 174102, doi:10.1103/PhysRevLett.93.174102.
7. Abrams, D.M.; Mirollo, R.; Strogatz, S.H.; Wiley, D.A. Solvable model for chimera states of coupled oscillators. *Phys. Rev. Lett.* **2008**, *101*, 084103, doi:10.1103/PhysRevLett.101.084103.
8. Tsigkri-DeSmedt, N.D.; Hizanidis, J.; Hövel, P.; Provata, A. Multi-chimera states in the Leaky Integrate-and-Fire model. *Procedia Comput. Sci.* **2015**, *66*, 13–22, doi:10.1016/J.PROCS.2015.11.004.
9. Omelchenko, I.; Omel'chenko, O.E.; Hövel, P.; Schöll, E. When nonlocal coupling between oscillators becomes stronger: Patched synchrony or multichimera states. *Phys. Rev. Lett.* **2013**, *110*, 224101, doi:10.1103/PhysRevLett.110.224101.
10. Hizanidis, J.; Kanas, V.G.; Bezerianos, A.; Bountis, T. Chimera states in networks of nonlocally coupled Hindmarsh–Rose neuron models. *Int. J. Bifurc. Chaos* **2014**, *24*, 1450030, doi:10.1142/S0218127414500308.
11. Kemeth, F.P.; Haugland, S.W.; Schmidt, L.; Kevrekidis, I.G.; Krischer, K.A. A classification scheme for chimera states. *Chaos Interdiscip. J. Nonlinear Sci.* **2016**, *26*, 094815, doi:10.1063/1.4959804.
12. Wolfrum, M.; Omel'chenko, O.E.; Yanchuk, S.; Maistrenko, Y.L. Spectral properties of chimera states. *Chaos Interdiscip. J. Nonlinear Sci.* **2011**, *21*, 013112, doi:10.1063/1.3563579.
13. Bogomolov, S.A.; Strelkova, G.I.; Schöll, E.; Anishchenko, V.S. Amplitude and phase chimeras in an ensemble of chaotic oscillators. *Tech. Phys. Lett.* **2016**, *42*, 765–768, doi:10.1134/S1063785016070191.
14. Panaggio, M.J.; Abrams, D.M. Chimera states: Coexistence of coherence and incoherence in networks of coupled oscillators. *Nonlinearity* **2015**, *28*, R67–R87, doi:10.1088/0951-7715/28/3/R67.
15. Pecora, L.M.; Sorrentino, F.; Hagerstrom, A.M.; Murphy, T.E.; Roy, R. Cluster synchronization and isolated desynchronization in complex networks with symmetries. *Nat. Commun.* **2014**, *5*, 4079, doi:10.1038/ncomms5079.
16. Kuznetsov, A.P.; Kuznetsov, S.P.; Shchegoleva, N.A.; Stankevich, N.V. Dynamics of coupled generators of quasiperiodic oscillations: Different types of synchronization and other phenomena. *Phys. D Nonlinear Phenom.* **2019**, *398*, 1–12, doi:10.1016/J.PHYSD.2019.05.014.
17. Kuznetsov, A.P.; Sataev, I.R.; Tyuryukina, L.V. Synchronization of quasi-periodic oscillations in coupled phase oscillators. *Tech. Phys. Lett.* **2010**, *36*, 478–481, doi:10.1134/S1063785010050263.
18. Anishchenko, V.; Nikolaev, S.; Kurths, J. Winding number locking on a two-dimensional torus: Synchronization of quasiperiodic motions. *Phys. Rev. E* **2006**, *73*, 056202, doi:10.1103/PhysRevE.73.056202.
19. Anishchenko, V.; Nikolaev, S.; Kurths, J. Peculiarities of synchronization of a resonant limit cycle on a two-dimensional torus. *Phys. Rev. E* **2007**, *76*, 046216, doi:10.1103/PhysRevE.76.046216.
20. Loose, A.; Wünsche, H.J.; Henneberger, F. Synchronization of quasiperiodic oscillations to a periodic force studied with semiconductor lasers. *Phys. Rev. E* **2010**, *82*, 035201, doi:10.1103/PhysRevE.82.035201.
21. Stankevich, N.V.; Kurths, J.; Kuznetsov, A.P. Forced synchronization of quasiperiodic oscillations. *Commun. Nonlinear Sci. Numer. Simul.* **2015**, *20*, 316–323, doi:10.1016/J.CNSNS.2014.04.020.
22. Clerc, M.G.; Coulibaly, S.; Ferré, M.A.; García-Ñustes, M.A.; Rojas, R.G. Chimera-type states induced by local coupling. *Phys. Rev. E* **2016**, *93*, 052204, doi:10.1103/PhysRevE.93.052204.
23. Gupte, N.; Singha, J. Classification and Analysis of Chimera States. In *Proceedings of the 5th International Conference on Applications in Nonlinear Dynamics*; Maui, Hawaii, USA, Aug. 5-9, 2018; Springer: Cham, Switzerland, 2019; pp. 318–328.

24. Velichko, A.; Belyaev, M.; Putrolaynen, V.; Perminov, V.; Pergament, A. Thermal coupling and effect of subharmonic synchronization in a system of two VO2 based oscillators. *Solid State Electron.* **2018**, *141*, 40–49, doi:10.1016/J.SSE.2017.12.003.
25. Velichko, A.; Belyaev, M.; Putrolaynen, V.; Boriskov, P.; Velichko, A.; Belyaev, M.; Putrolaynen, V.; Boriskov, P. A new method of the pattern storage and recognition in oscillatory neural networks based on resistive switches. *Electronics* **2018**, *7*, 266, doi:10.3390/electronics7100266.
26. Velichko, A.; Belyaev, M.; Boriskov, P. A model of an oscillatory neural network with multilevel neurons for pattern recognition and computing. *Electronics* **2019**, *8*, 75, doi:10.3390/electronics8010075.
27. Pergament, A.L.; Boriskov, P.P.; Velichko, A.A.; Kuldin, N.A. Switching effect and the metal–insulator transition in electric field. *J. Phys. Chem. Solids* **2010**, *71*, 874–879, doi:10.1016/J.JPCS.2010.03.032.
28. Belyaev, M.A.; Boriskov, P.P.; Velichko, A.A.; Pergament, A.L.; Putrolainen, V.V.; Ryabokon', D.V.; Stefanovich, G.B.; Sysun, V.I.; Khanin, S.D. Switching channel development dynamics in planar structures on the basis of Vanadium Dioxide. *Phys. Solid State* **2018**, *60*, 447–456, doi:10.1134/S1063783418030046.
29. Joushaghani, A.; Jeong, J.; Paradis, S.; Alain, D.; Stewart Aitchison, J.; Poon, J.K.S. Voltage-controlled switching and thermal effects in VO$_2$ nano-gap junctions. *Appl. Phys. Lett.* **2014**, *104*, 221904, doi:10.1063/1.4881155.
30. Itoh, M.; Chua, L.O. Star cellular neural networks for associative and dynamic memories. *Int. J. Bifurc. Chaos* **2004**, *14*, 1725–1772, doi:10.1142/S0218127404010308.
31. Velichko, A.; Belyaev, M.; Putrolaynen, V.; Boriskov, P.; Velichko, A.; Belyaev, M.; Putrolaynen, V.; Boriskov, P. A new method of the pattern storage and recognition in oscillatory neural networks based on resistive switches. *Electronics* **2018**, *7*, 266, doi:10.3390/electronics7100266.
32. Jerry, M.; Ni, K.; Parihar, A.; Raychowdhury, A.; Datta, S. Stochastic insulator-to-metal phase transition-based true random number generator. *IEEE Electron Device Lett.* **2018**, *39*, 139–142, doi:10.1109/LED.2017.2771812.
33. Velichko, A.; Belyaev, M.; Putrolaynen, V.; Perminov, V.; Pergament, A. Modeling of thermal coupling in VO$_2$ based oscillatory neural networks. *Solid State Electron.* **2018**, *139*, 8–14, doi:10.1016/j.sse.2017.09.014.
34. Perminov, V.V.; Putrolaynen, V.V.; Belyaev, M.A.; Velichko, A.A. Synchronization in the system of coupled oscillators based on VO$_2$ switches. *J. Phys. Conf. Ser.* **2017**, *929*, 012045, doi:10.1088/1742-6596/929/1/012045.
35. Yao, X.S.; Maleki, L.; Davis, L. Coupled opto-electronic oscillators. In Proceedings of the 1998 IEEE International Frequency Control Symposium (Cat. No.98CH36165), Pasadena, CA, USA, 29 May 1998; pp. 540–544.
36. Kravtsov, K.S.; Fok, M.P.; Prucnal, P.R.; Rosenbluth, D. Ultrafast all-optical implementation of a leaky integrate-and-fire neuron. *Opt. Express* **2011**, *19*, 2133, doi:10.1364/OE.19.002133.
37. Zong, Y.; Dai, X.; Gao, Z.; Busawon, K.; Binns, R.; Elliott, I. Synchronization of pulse-coupled oscillators for IEEE 802.15.4 Multi-Hop wireless sensor networks. In Proceedings of the *2018 IEEE Global* Communications Conference *(GLOBECOM)*, Abu Dhabi, UAE, 9–13 December 2018; pp. 1–7.
38. Behera, S.S.; Chattopadhyay, S.A. Comparative Study of back propagation and simulated annealing algorithms for neural net classifier optimization. *Procedia Eng.* **2012**, *38*, 448–455, doi:10.1016/J.PROENG.2012.06.055.
39. Zhang, J.; Gao, Y. Synchronization of coupled neural networks with time-varying delay. *Neurocomputing* **2017**, *219*, 154–162, doi:10.1016/J.NEUCOM.2016.09.004.
40. Shang, Y. On the delayed scaled consensus problems. *Appl. Sci.* **2017**, *7*, 713, doi:10.3390/app7070713.
41. Velichko, A.A.; Belyaev, M.A. An Investigation of the Effect of the Thermal Coupling Time Delay on the Synchronization of VO2-Oscillators. *Tech. Phys. Lett.* **2019**, *45*, 61–64, doi:10.1134/S1063785019020184.
42. Shang, Y. Synchronization in networks of coupled harmonic oscillators with stochastic perturbation and time delays. *Ann. Acad. Rom. Sci. Ser. Math. Appl.* **2012**, *4*, 44–58.
43. Shang, Y. Resilient multiscale coordination control against adversarial nodes. *Energies* **2018**, *11*, 1844, doi:10.3390/en11071844.
44. Shang, Y. Resilient consensus of switched multi-agent systems. *Syst. Control. Lett.* **2018**, *122*, 12–18, doi:10.1016/J.SYSCONLE.2018.10.001.
45. Li, C.; Li, Y. Fast and robust image segmentation by small-world neural oscillator networks. *Cogn. Neurodyn.* **2011**, *5*, 209–220, doi:10.1007/s11571-011-9152-2.

46. Vodenicarevic, D.; Locatelli, N.; Abreu Araujo, F.; Grollier, J.; Querlioz, D.A Nanotechnology-Ready computing scheme based on a weakly coupled oscillator network. *Sci. Rep.* **2017**, *7*, 44772, doi:10.1038/srep44772.
47. Pikovsky, A.; Rosenblum, M. Self-Organized partially synchronous dynamics in populations of nonlinearly coupled oscillators. *Phys. D Nonlinear Phenom.* **2009**, *238*, 27–37, doi:10.1016/J.PHYSD.2008.08.018.
48. Xu, C.; Sun, Y.; Gao, J.; Qiu, T.; Zheng, Z.; Guan, S. Synchronization of phase oscillators with frequency-weighted coupling. *Sci. Rep.* **2016**, *6*, 21926, doi:10.1038/srep21926.
49. Shang, Y. An agent based model for opinion dynamics with random confidence threshold. *Commun. Nonlinear Sci. Numer. Simul.* **2014**, *19*, 3766–3777, doi:10.1016/J.CNSNS.2014.03.033.
50. Shang, Y. Deffuant model with general opinion distributions: First impression and critical confidence bound. *Complexity* **2013**, *19*, 38–49, doi:10.1002/cplx.21465.
51. Velichko, A.; Putrolaynen, V.; Belyaev, M. Effects of higher order and long-range synchronizations for classification and computing in oscillator-based spiking neural networks. *arXiv* **2018**, arXiv:1804.03395.